\documentstyle[twoside]{article}
\def\SU{{\rm SU}}
\def\SO{{\rm SO}}
\def\Tr{\mathop{\rm Tr}}
\def\diag{\mathop{\rm diag}}
\newcommand{\eu}{\varepsilon_u}
\newcommand{\ed}{\varepsilon_d}
\newcommand{\md}[1]{\langle #1\rangle}

\makeatletter
\newcounter{alphaequation}[equation]
\def\thealphaequation{\theequation\alph{alphaequation}}
\def\eqnsystem#1{
\def\@eqnnum{{\rm (\thealphaequation)}}
\def\@@eqncr{\let\@tempa\relax
\ifcase\@eqcnt \def\@tempa{& & &}
\or \def\@tempa{& &}\or \def\@tempa{&}\fi\@tempa
\if@eqnsw\@eqnnum\refstepcounter{alphaequation}\fi
\global\@eqnswtrue\global\@eqcnt=0\cr}
\refstepcounter{equation}
\let\@currentlabel\theequation
\def\@tempb{#1}
\ifx\@tempb\empty\else\label{#1}\fi
\refstepcounter{alphaequation}
\let\@currentlabel\thealphaequation
\global\@eqnswtrue\global\@eqcnt=0
\tabskip\@centering\let\\=\@eqncr
$$\halign to \displaywidth\bgroup
  \@eqnsel\hskip\@centering
  $\displaystyle\tabskip\z@{##}$&\global\@eqcnt\@ne
  \hskip2\arraycolsep\hfil${##}$\hfil&
  \global\@eqcnt\tw@\hskip2\arraycolsep
  $\displaystyle\tabskip\z@{##}$\hfil
  \tabskip\@centering&\llap{##}\tabskip\z@\cr}

\def\endeqnsystem{\@@eqncr\egroup$$\global\@ignoretrue}
\makeatother

\begin{document}\large
\hfill\vbox{\baselineskip12pt
            \hbox{\bf IFUP -- TH 32/94}
            \hbox{\bf  hep-ph/9407239}
            \hbox{June 1994}}
\vspace{7mm}
\begin{center}\vglue 0.6cm{\LARGE\bf\vglue 10pt
   Fermion masses and mixings\\ \vglue 3pt
   in a flavour symmetric GUT    \\}
\vglue 1.0cm
{\large\bf  Riccardo Barbieri, Gia Dvali$^*$ and Alessandro Strumia\\[4mm] }
\baselineskip=13pt
{\em Dipartimento di Fisica, Universit\`a di Pisa}\\[1mm]
{\rm and}\\[1mm]
{\em INFN, Sezione di Pisa, I-56126 Pisa, Italy}\\
\vfill

{\large\bf Abstract}
\end{center}

\vglue 0.3cm{\rightskip=3pc \leftskip=3pc \tenrm\baselineskip=12pt
\noindent\large
We describe a supersymmetric Grand Unified Theory based on the gauge group
$\SU(5)^3$, or $\SO(10)^3$, invariant under the interchange of any~SU(5),
or~SO(10), with each family multiplet transforming non trivially under
one different individual group factor.
A realistic pattern of fermion masses and mixings is obtained as a result
of an appropriate choice out of the many possible discretely degenerate
vacua of the theory.\\ \indent
In the~SO(10) case, we predict the three neutrino masses in terms of an
overall scale and, within factors of order unity, their mixing angles.
A $\nu_e$-$\nu_\mu$ oscillation is suggested as a solution of the solar
neutrino problem, implying a visible $\nu_\mu$-$\nu_\tau$ oscillation
in the forthcoming experiments.\\ \indent
Grand unified theories of this type could be obtainable in a
string theory framework.}

\vfill~\vfill\normalsize\footnoterule~\\[0.5mm]
\noindent
{$\ast$~~Permanent address:\em{}
Institute of Physics, Georgian Academy of Sciences, 380077 Tbilisi, Georgia.}

\thispagestyle{empty}
\newpage~
\setcounter{page}{1}

\normalsize
\section{Introduction}
Finding a rationale in the observed pattern of fermion masses
and mixings is one of the most challenging problems in today particle physics.
Certainly in the Standard Model, but also in current GUTs, it is in describing
the flavour sector that most of the elegance of the gauge theory
structure gets lost in a plethora of arbitrary parameters.
Perhaps most annoying than everything else is the fact that
the clear hierarchical pattern of fermion masses is accounted for by a
progression of smaller and smaller Yukawa couplings.
In the language of symmetries, when the Yukawa couplings are
switched off, the Standard Model lagrangian has a huge unobserved global
symmetry which must be broken in a hierarchical way.
Current GUTs, by extending the gauge group, do reduce such a global
symmetry but always leave intact the part of it which is related
to the family replicas.
The fate of this residual symmetry is, to a large extent, the hearth
of the flavour problem.

In this work we propose a supersymmetric GUT based on the gauge group
$\SU(5)^3$, or~$\SO(10)^3$,
which leaves no residual non-abelian flavour symmetry
and where, in fact, flavour is treated in a completely symmetric way.
The triplication of the~SU(5) (SO(10)) factors of the overall gauge group is
in one-to-one correspondence with the three family replicas.
A realistic pattern of fermion masses and mixings is obtained
as a result of an appropriate choice out of the many possible
discretely degenerate vacua of the theory.
First of all, such a choice determines the lightness of a pair of
Higgs doublets, which are split from the coloured triplet partners.
In turn, this is where the seed of the
hierarchy in the fermion masses resides.
The light Higgs doublets do transform only under one of the
individual~SU(5) factors of the gauge group.
As a consequence, only the fermions of one family have a sizeable
coupling to the light Higgs bosons, whereas the lighter families get coupled
to them by a stepwise procedure,
again determined by a specific vacuum structure.
As a result, it is possible to describe the quark and lepton masses
and mixings in terms of two/three appropriately chosen small parameters,
a part from dimensionless couplings of order unity.
Along these lines, by going to~SO(10), we predict the three neutrino
masses in term of an overall scale and, within factors of order
unity, their mixing angles.
A $\nu_e$-$\nu_\mu$ oscillation is suggested as a solution of the solar
neutrino problem, implying a visible $\nu_\mu$-$\nu_\tau$ oscillation
in the forthcoming experiments at {\sc CERN}.

As an independent motivation for such kind of theories, we are lead
to consider models based on gauge groups made of several identical group
factors (two or more) by the considerations of unified theories whose
symmetry can be reduced to the standard $\SU(3)\otimes\SU(2)\otimes{\rm U}(1)$
by Higgs fields in the fundamental representation~\cite{GUTAndString}.
There are reasons to think that such theories can be obtained in a
string theory framework~\cite{StringAndGUT}.
Since the triplication of the gauge group factor is related to the
number of families, it appears that~$\SO(10)^3$
is the largest possible gauge group that one can obtain starting from
the $\SO(32)$ symmetry of string theory~\cite{Strings}.

\section{Symmetries and multiplet structure}
The gauge group is
\begin{equation}
\SU(5)_1\otimes\SU(5)_2\otimes\SU(5)_3=\bigotimes_{i=1}^3\SU(5)_i
\end{equation}
The index $i$ is associated with the family index in the sense that the
$i$-th family multiplet
\begin{equation}
f_i=(\bar{5}\oplus10)_i,\qquad i = 1,2,3
\end{equation}
only transforms under the $i$-th factor of the gauge group\footnote{Different
family multiplets transforming under different gauge group factors have been
advocated by Froggatt, Lowe and Nielsen~\cite{Froggatt}.
Our approach differs from theirs both because we work in a unified
context and because we require permutation invariance between different
families.}. We require invariance of the theory under the permutations
${\cal P}_{ij}$ of any pair of two indices $i,j=1,2,3$.

The minimal set of fields needed to break the gauge group down to standard
$\SU(3)\otimes\SU(2)\otimes{\rm U}(1)$ consistently with permutation
symmetry is given by the triplet
\begin{equation}
Z^{a_i}_{a_k}\equiv Z_{ik},\qquad
\bar{Z}^{a_k}_{a_i}\equiv \bar{Z}_{ik},\qquad
i<k;~i,k=1,2,3
\end{equation}
where the lower (upper) indices $a_i$ are in the (anti-)fundamental
representation of the $i$-th~SU(5).
With the definition
$$\bar{Z}_{ik}\equiv Z_{ki}$$
the permutation symmetry acts on the $Z$-fields by simple interchange
of the corresponding indices.
Finally the light Higgs doublets are contained in
\begin{equation}
H_{a_i}\equiv H_i,\qquad \bar{H}^{a_i}\equiv \bar{H}_i,\qquad i = 1,2,3,
\end{equation}
with an obvious action on them of ${\cal P}_{ij}$.

\section{Higgs superpotential}
Let us define (recalling that $i<k$)
\begin{equation}
\begin{array}{ll}
Z_{ik}\equiv Z_j & j\neq i,~j\neq k\\
\bar{Z}_{ik}\equiv \bar{Z}_j & j\neq i,~j\neq k\\
\end{array}
\end{equation}
The most general superpotential, $W(Z)$, dependent upon the $Z$-fields,
up to quartic terms has the form
\begin{eqnarray}\label{eq:WZ}
W(Z) &=& M_1\sum_j(Z_j\bar{Z}_j)+
\frac{1}{M_2}\sum_{j\neq k}(Z_j\bar{Z}_j)(Z_k\bar{Z}_k)+
\frac{1}{M_3}\sum_j(Z_j\bar{Z}_j)^2+\\
&&+
\frac{1}{M_4}\sum_{j\neq k}(Z_j\bar{Z}_j Z_k \bar{Z}_k)+
\frac{1}{M_5}\sum_j(Z_j\bar{Z}_j Z_j\bar{Z}_j)\nonumber
\end{eqnarray}
where the parentheses denote traces over the gauge group indices.
For simplicity we impose invariance of $W(Z)$ under $Z\to-Z$, even though
the inclusion of cubic terms would not alter the final result,
and we do not consider higher order terms.
We look for a
supersymmetric minimum of the potential of the form
\begin{equation}
(Z_j)^a_b = z_{j a} \delta^a_b,\qquad
(\bar{Z}_j)^a_b = \bar{z}_{ja} \delta^a_b
\end{equation}
The condition of vanishing of the $F$-terms
\begin{equation}
{\partial W \over \partial Z_j} = {\partial W \over \partial \bar{Z}_j}=0
\end{equation}
as well as of the $D$-terms gives rise to a large number of discretely
degenerate minima.
Of physical interest are the vacuum configurations of the form,
for a given $j$,
\begin{eqnsystem}{vacua}
\md{Z_j}=\md{\bar{Z_j}}  &=&  T_j\cdot\diag(1,1,1,0,0)\label{eq:VacT}\\
\md{Z_j}=\md{\bar{Z_j}}  &=&  D_j\cdot\diag(0,0,0,1,1)\label{eq:VacD}\\
\md{Z_j}=\md{\bar{Z_j}}  &=&  V_j\cdot\diag(1,1,1,a,a),\label{eq:VacTD}
\qquad a\neq0.\end{eqnsystem}
The main constraint, in a given triplet representation
of the permutation group, is that, if two or more fields have the same
configuration, then the modulus of their vacuum expectation values coincide.
It is only with the inclusion of higher order terms
in the superpotential~(\ref{eq:WZ})
that these vacuum expectation values can, but need not, be split.

Examples of minima which give rise to the desired breaking of the
gauge group down to the standard
$\SU(3)\otimes\SU(2)\otimes{\rm U}(1)$ are
(with a choice of the permutation indices),
\begin{equation}\label{eq:VacVT}
Z_{12}=Z_{13}=V\cdot\diag(1,1,1,a,a),\qquad
Z_{23}=T\cdot\diag(1,1,1,0,0),
\end{equation}
or
\begin{equation}\label{eq:vacVDT}
Z_{12}=V\cdot\diag(1,1,1,a,a),\qquad
Z_{13}=D\cdot\diag(0,0,0,1,1),\qquad
Z_{23}=T\cdot\diag(1,1,1,0,0).
\end{equation}
Around these minima, no state remains light other than the eaten up
Goldstone bosons.

Out of the $H$-fields, one wants a pair of~SU(2) doublets to remain light.
This is achieved by imposing a ${\cal Z}_2$
symmetry on the $H$-fields and on the $Z$-triplet of fields that are
coupled to them in the superpotential, under which
\begin{equation}\label{eq:Z2}
\{\bar{H},Z,\bar{Z}\} \to - \{\bar{H},Z,\bar{Z}\}
\end{equation}
whereas $H$ stays invariant.
At renormalizable level, the superpotential coupling the $H$ with the
$Z$-fields has the form
\begin{equation}\label{eq:WHZ}
W(H,Z) = \lambda\sum_{i,k} H_i Z_{ik} \bar{H}_k,
\end{equation}
or, more explicitly,
\begin{equation}
W(H,Z)=\lambda \matrix{ (\bar{H}_1,\bar{H}_2,\bar{H}_3) \cr~ \cr~}
\pmatrix{0&Z_{12}&Z_{13}\cr
\bar{Z}_{12}&0&Z_{23}\cr
\bar{Z}_{13}&\bar{Z}_{23}&0}
\pmatrix{H_1\cr H_2\cr H_3}.
\end{equation}
As a possible example,
if the $Z$-triplet entering into $W(Z)$, eq.~(\ref{eq:WZ}), and
$W(H,Z)$, eq.~(\ref{eq:WHZ}), coincide,
in the vacuum~(\ref{eq:VacVT}) all the $H$-multiplets get a mass except
a pair of SU(2)-doublets.
Even with the inclusion in the superpotential of higher dimensional terms,
these doublets remain light provided, the operators of factorized form
\begin{equation}\label{eq:Sopprimenda}
(H Z^n\bar{H})\Tr (Z^{m+1}),\qquad n+m=2k,\qquad k=1,2,\ldots,
\end{equation}
are sufficiently suppressed.
In this vacuum configuration, with $V$ and $T$ both of order $10^{16}\,$GeV,
the observed strong and electroweak coupling constants unify at a common
value $g/\sqrt{3}$, where $g$ is the original gauge coupling.

\section{Yukawa superpotential and fermion mass matrices. General expressions}
The Yukawa superpotential has the following generic form
\begin{equation}\label{eq:WY}
W_Y =
\lambda^{\rm u} \sum_i H_i\,10_i~10_i+
\lambda^{\rm d} \sum_i\bar{H}_i\,10_i~\bar{5}_i+
\sum_{n,m}\frac{\lambda^{\rm u}_{nm}}{M_u^{n+m}} H Z^n~10~Z^m~10+
\sum_{n',m'}\frac{\lambda^{\rm d}_{nm}}{M_d^{n+m}}
\bar{H} Z^{n'}~10~Z^{m'}~\bar{5}
\end{equation}
where the higher dimensional operators have been written in schematic form.
We have in mind the possibility that the higher dimensional operators
in~(\ref{eq:WY}) be generated by exchanges of heavy vectorlike states,
which might give rise to different energy scales relevant to the
$H~10~10$ and to the $\bar{H}~10~\bar{5}$ operators,
$M_u$ and $M_d$ respectively.
Operators possibly contributing to neutrino masses will be discussed
later on, in a more interesting~SO(10) context.

If a unique $Z$-triplet enters $W(Z)$ and $W(H,Z)$,
and if we take~(\ref{eq:VacVT}) as the relevant vacuum configuration,
the light Higgs doublets would be found in the linear combinations
$(H_2-H_3)/\sqrt{2}$ and $(\bar{H}_2-\bar{H}_3)/\sqrt{2}$.
As a consequence, from the renormalizable terms in $W_Y$, one family
remains massless whereas the two others are degenerate in the~10 and
in the $\bar{5}$ sectors separately.
Assuming a small perturbation from higher order terms, although we do see
a possible connection between the lightness of the Higgs doublets and
the hierarchical structure of the fermion masses, this is not a
reasonable starting point.
To cure it, one can either assume a significant splitting between the
lower components of $Z_{12}$ and $Z_{13}$ in the vacuum~(\ref{eq:VacVT}),
or take a different $Z$-triplet entering $W(H,Z)$ with only one
component, say $Z_{12}$, having a non vanishing vacuum expectation value
in the two lower entries.
In this last case, to which we stick,
the light doublets reside in $H_3,\bar{H}_3$.

{}From $W_Y$, the insertion of the $Z$-vacuum expectation value
in the upper three or
lower two components, called $T$ and $D$ respectively hereafter,
generates the effective Yukawa couplings to the light doublets.
The higher dimensional operators play an essential role both in generating
the hierarchy of the diagonal couplings, since the light doublets
only reside in the $H_3$ ($\bar{H}_3$), and in generating the off-diagonal
entries.
At the $\SU(3)\otimes\SU(2)\otimes{\rm U}(1)$ level, the structure
of the higher order operators for the up and down mass matrices are the
following
\begin{equation}
H D^n Q T^m u^c,\qquad
\bar{H} D^{n'} Q T^{m'} d^c
\end{equation}
whereas, for the leptons, one has
\begin{equation}
\bar{H} D^{n''} L D^{m''} e^c
\end{equation}
$T$'s are essential to give rise to off-diagonal quark operators,
whereas for leptons off-diagonal operators are only generated by
vacuum expectation values in the $D$-direction.
The different fermion mass matrices,
\begin{equation}\label{eq:MMM}
Q_i M^{\rm u}_{ij} u_j^c,\qquad
Q_i M^{\rm d}_{ij} d_j^c,\qquad
L_i M^{\rm e}_{ij} e_j^c,\qquad
\end{equation}
by making an expansion in $T/M_{u,d}$, $D/M_{u,d}$, have therefore
the following general form
\begin{equation}
M_{ij}^{\rm u,d,e} = \mu_i^{\rm u,d,e} V_{ij}^{\rm u,d,e}
\end{equation}
where (with the usual meaning of the angle $\beta$),
\begin{eqnarray}
\mu_i^{\rm u} &=& v\sin\beta\left(\lambda^{\rm u} \delta_{3i} +
\frac{D_{3i}}{M_u} + \sum_j \frac{D_{3j}D_{ji}}{M_u^2}+\cdots\right)\\
\mu_i^{\rm d,e} &=& v\cos\beta\left(\lambda^{\rm d} \delta_{3i} +
\frac{D_{3i}}{M_d} + \sum_j \frac{D_{3j}D_{ji}}{M_d^2}+\cdots\right),
\end{eqnarray}
and
\begin{eqnsystem}{sys:Vude}
V_{ij}^{\rm u} &=& \delta_{ij} +\frac{(T_{ij})^2}{M_u^2}+\cdots\\
V_{ij}^{\rm d} &=& \delta_{ij} +\frac{T_{ij}}{M_d}+\sum_k
\frac{T_{ik} T_{kj}}{M_d^2}+\cdots\\
V_{ij}^{\rm e} &=& \delta_{ij} +\frac{D_{ij}}{M_d}+\sum_k
\frac{D_{ik} D_{kj}}{M_d^2}+\cdots
\end{eqnsystem}
In front of every mass term involving a $D$ or a $T$, a dimensionless
coupling has been left understood, in general also different for
leptons and down quarks.
Possible relations between the corresponding entries in the $d$ and the
$\ell$-mass matrices may in fact occur, as a remnant of the SU(5) symmetry,
depending on the more specific structure of the theory (see below).
The difference in the~SU(5) representations that contain the $u^c$ (10)
and the $d^c$ ($\bar{5}$) is the source of the difference in the\
corresponding off-diagonal terms in $V_{ij}^{\rm u}$ and $V_{ij}^{\rm d}$.

It is manifest from the structure of the mass
matrices~(\ref{eq:MMM}--\ref{sys:Vude}) that
$D$ and $T$-vacuum expectation values smaller than $M_{u,d}$ can induce
a hierarchy in the masses and in the mixing angles of the fermions,
in spite of the original flavour symmetry.
As we shall see, this possibility has everything to do with the
huge number of different vacua present in the theory.
The degeneracy of those vacua will be lifted as a consequence of
supersymmetry breaking and the vacuum of interest will be converted
into a global or a local minimum depending upon the specific form
of the potential and its parameters.

At the same time it is important to observe, as a neat property of the
theory that we are discussing, that the tree level soft supersymmetry
breaking masses of the squarks (and of the leptons) are necessarily
flavour degenerate if supersymmetry breaking, as expected,
preserves the permutation symmetry.

\section{Fermion masses matrices: towards a realistic structure}
Other than its interesting features, however, the general kind of theories
that have been discussed so far have two quite manifest problems.
The first one is the lack of sufficient asymmetry between the up and
down quark mass matrices, which is especially embarrassing in view of the
large top-bottom mass difference.
The second one is related to the proton decay amplitude mediated by triplet
exchanges.
In fact, we generally expect such amplitude to be significantly larger than
in a standard ``minimal'' SU(5) theory, since at least one of the~3
coloured triplets in $H_i$ will have large couplings, of order unity,
to the first generation as a consequence of the permutation symmetry.
Even if the mass of such triplet were pushed to the Planck scale,
the corresponding amplitude could not be sufficiently suppressed.

We think that the only common solution of both problems resides in
barring the renormalizable coupling of the down-type quarks in the Yukawa
superpotential, as it can be enforced by an appropriate set of
${\cal Z}_2$-symmetries, which extend the one defined in eq.~(\ref{eq:Z2}).
In particular, we propose the following structure as a possible
example of a realistic theory of the mass matrices.

The $H$-fields occur in two replicas
\begin{equation}
H_i,\bar{H_i};\qquad H'_i,\bar{H}'_i,
\end{equation}
whereas the $Z$-fields occur in several replicas, which must include
the fields
\begin{equation}
Z,Z',Y,Y'.
\end{equation}
On these fields it is simple to write down a set of ${\cal Z}_2$-symmetries
which enforce the following structure of the Higgs and Yukawa superpotential
(omitting dimensionless couplings and summations over permutation indices)
\begin{eqnarray}\label{eq:WHZ2}
W(H,Z) &=& HZ\bar{H}+H'Z'\bar{H}'\\
W_Y &=& H~10~10 + H\, Y^n~10~Y^m~10+\bar{H}'\,Y'~10~\bar{5}+
\bar{H}'\,Y'~Y^{n'}~10~Y^{m'}\,\bar{5}.
\end{eqnarray}
With these superpotentials, out of the various possibilities,
a vacuum which leads to a realistic structure of the mass matrices
is the one where the fields that have a non-vanishing vacuum expectation
value in the $D$-direction are $Z_{12}$, $Z'_{13}$, $Y_{12}$, $Y_{23}$
and $Y'_{23}$. Furthermore all these fields have non-zero
vacuum expectation values in the $T$-direction except for the $Y'$.
The essential properties of this choice are:
\begin{itemize}
\item $D_{Z_{12}}\neq0$ only (out of the $Z_{ik}$'s)
forces the light doublet coupled to the up-quarks
to reside in the $H_3$ direction only.
\item In the same way $D_{Z'_{13}}\neq0$ gives the light doublet coupled
to the down-quarks (and the leptons) in the $H_2$ direction.
However, since $D_{Y'_{23}}\neq0$ only, $H_2$ has its main effective coupling,
although suppressed by a factor
\begin{equation}
\varepsilon'\equiv{\md{D_{Y'_{23}}}\over M_d}
\end{equation}
to the third generation.
\item The vanishing of $D_{Y_{13}}$ is the source of the hierarchy in the
eigenvalues of the mass matrices, suppressed relative to the dominant
third one by factors $\varepsilon_d,\varepsilon_u,\varepsilon_d^2,
\varepsilon_u^2$ where
\begin{equation}
\varepsilon_d\equiv{\md{D_{Y_{13}}}\over M_d},\qquad
\varepsilon_u\equiv{\md{D_{Y_{13}}}\over M_u}.
\end{equation}
\item The vanishing of all the $Y'$ fields in the $T$-direction is the
source of the decoupling of the $d$-type quarks from the heavy triplets
in $\bar{H}$~\cite{Dvali}, whose exchange would otherwise
lead to the proton decay at a dangerous level.
This last property requires some suppression
of the operators $(\bar{H}'\,10~\bar{5})\Tr(Y'Y^n)$.
Notice that also these operators,
as it was the case for~(\ref{eq:Sopprimenda})
are of factorized form in the gauge group indices.
On this basis, from now on, we shall assume that all factorized operators
are absent or have anyhow a negligible strength.
Notice also that the Higgs superpotential~(\ref{eq:WHZ2}) leads, as it
stands, to two pairs of light doublets, which is undesiderable from the
point of view of the unification of the coupling constants.
A possible way out consists in introducing the coupling
$S\bar{H}H'$, again consistently with the ${\cal Z}_2$-symmetries,
with the singlet $S$ getting a vacuum expectation value at a scale
greater than or equal to the unification scale.
SO(10) offers an alternative, more interesting, solution for
this doubling of the light Higgs pairs (see below).
\end{itemize}
A systematic exploration of the Yukawa superpotential $W_Y$ with the
assumed properties of the relevant vacuum leads to the following
structure of the mass matrices~(\ref{eq:MMM})
(leaving understood a numerical coupling, $\lambda_{ij}^{\rm u,d,e}$,
of order unity in front of every entry)
\begin{eqnsystem}{sys:ude}
M^{\rm u}  &=& v\sin\beta \pmatrix{
\eu^2     & \eu^2 t_u^2 & \eu^2 t_u^2 \cr
\eu t_u^2 & \eu         & \eu   t_u^2 \cr
    t_u^2 &       t_u^2 &1 }\\
M^{\rm d}  &=& v\varepsilon'\cos\beta\pmatrix{
\ed^2     & \ed^2 t_d & \ed^2 t_d \cr
\ed   t_d & \ed       & \ed   t_d \cr
      t_d &       t_d &1 }\\
M^{\rm e}  &=& v\varepsilon'\cos\beta\pmatrix{
\ed^2 & \ed^2 &\ed^2\cr
\ed^3 & \ed   & \ed\cr
\ed^2 & 1     & 1}
\end{eqnsystem}
where
\begin{equation}
t_u = \frac{\md{T_Y}}{M_u},\qquad t_d = \frac{\md{T_Y}}{M_d}.
\end{equation}
Some relations occur between the numerical couplings $\lambda^{\rm d}_{ij}$
and $\lambda^{\rm e}_{ij}$ in the $d$ and $\ell$-mass matrices,
as a consequence of the SU(5) symmetry.
An equality in fact holds, but only for the $33$ and $11$ entries.
This is because the $33$ entries only arise from the operator
(with an obvious contraction of the group indices)
\begin{equation}
\bar{H}'_2 \bar{Y}'_{23}~10_3~\bar{5}_3,
\end{equation}
and the $11$ entries come from
\begin{equation}
\bar{H}'_2 \bar{Y}'_{23}Y_{23} Y_{12}~10_1~\bar{5}_1.
\end{equation}
On the contrary, an operator which contributes at dominant level to the
$22$ entry of the leptons only is
\begin{equation}
\bar{H}'_2~10_2~Y_{23}Y'_{32}\bar{5}_2.
\end{equation}
As a further significant asymmetry between the two mass matrices,
notice also the operator
\begin{equation}
\bar{H}'_2~10_2~\bar{Y}_{32}'\bar{5}_3,
\end{equation}
which contributes, at dominant level, to the $23$ entry
for the leptons only ($Y'_{23}$ has no vacuum expectation value in the
$T$-direction).

The mass matrices~(\ref{sys:ude}) offer the possibility of a realistic
description of fermion masses and mixings in terms of the
parameters $\eu,\ed,t_d\,(t_u=t_d\eu/\ed)$ and $\varepsilon'$
(a part from couplings of order unity) which are related to
a hierarchy of scales.
Without attempting a fit, in view of the unknown coefficients, but rather
taking for illustration
\begin{equation}
\ed=\frac{1}{20},\qquad \eu = \frac{1}{400},
\end{equation}
one has, within the present uncertainties and after appropriate
renormalization group rescalings~\cite{RGE} at the unification scale
\begin{eqnarray*}
\frac{m_c}{m_t}  &=& (0.3\div2.0)\cdot\eu\\
\frac{m_u}{m_c}  &=& (1.0\div2.5)\cdot\eu\\
\frac{m_s}{m_b}  &=& (0.2\div1.0)\cdot\ed\\
\frac{m_d}{m_s}  &=& (0.5\div2.0)\cdot\ed\\
\frac{m_\mu}{m_\tau} &=& 1.20\cdot\ed\\
V_{us}             &=& 4.41\cdot\ed\\
\frac{m_e}{m_\mu}  &=& 0.097\cdot\ed
\end{eqnarray*}
Furthermore, with $\varepsilon'=\tan\beta/200$ and $t_d=1$,
\begin{eqnarray*}
\frac{m_b}{m_t}  &=& (0.2\div5)\cdot\frac{\varepsilon'}{\tan\beta} \\
\frac{V_{cb} m_b}{m_s}  &=& (0.5\div2.5)\cdot t_d\\
\frac{V_{ub}m_s}{V_{cb} m_b}  &=& (0.5\div3.0)\cdot t_d
\end{eqnarray*}
The greatest uncertainty in the various coefficients
is due to the poor knowledge
of the top Yukawa coupling at the unification scale, which we
take to vary from~0.3 to~4, so that $M_t^{\rm pole}=(150\div200)\,{\rm GeV}$
and the perturbative expansion is maintained up to unification.

In view of the unknown coefficients of order unity that enter
the matrices~(\ref{sys:ude}), we find an overall remarkably consistent
picture of all masses and mixings.
This is achieved with the choice of scales
\begin{equation}
M_u\approx 20\, \{M_d,T_Y\}\approx 400 D_Y \approx
\frac{4000}{\tan\beta} D_{Y'}
\end{equation}
Notice that, due to the smallness of the parameter $\eu$, the elements of
the CKM matrix are only determined by $M^{\rm d}$.

\section{Extension to SO(10)$^3$}
The model described in the previous sections can be extended to the gauge
group~$\SO(10)^3$ with two aspects deserving a special discussion:
the solution of the proton decay problem
(related to the up/down mass splitting) and the neutrino masses.
The Higgs fields required to break the group down to
$\SU(3)\otimes\SU(2)\otimes{\rm U}(1)$ are a triplet of fields with a pair of
vector indices in two of the three~SO(10) factors
\begin{equation}
Z_{a_i a_k}\equiv Z_{ik},\qquad i<k;\qquad i,k=1,2,3,
\end{equation}
and a triplet of fields transforming as the spinorial representation
of the same~SO(10)'s, together with the corresponding conjugate
representations
\begin{equation}
\psi_{\alpha_i}\equiv \psi_i;\qquad \bar\psi_{\alpha_i}\equiv\bar\psi_i,\qquad
i=1,2,3.
\end{equation}
{}From a generic superpotential $W(Z)$, the discussion of the vacuum
expectation values of the fields $Z_{ik}$ is identical to the one
of section~3 for the~SU(5) case.
Of interest are the vacuum configurations
\begin{eqnsystem}{vacua10}
\md{Z}=\md{\bar{Z}}&=&T_j\cdot\diag(1,1,1,1,1,1,0,0,0,0)\label{eq:VacT10}\\
\md{Z}=\md{\bar{Z}}&=&D_j\cdot\diag(0,0,0,0,0,0,1,1,1,1)\label{eq:VacD10}\\
\md{Z}=\md{\bar{Z}}&=&V_j\cdot\diag(1,1,1,1,1,1,a,a,a,a),\label{eq:VacTD10}
\qquad a\neq0.\end{eqnsystem}
which break the $\SO(10)\otimes\SO(10)$ group, under which the individual
$Z_{ik}$ trasform, down to
\begin{eqnarray*}
&&\SO(6)\otimes\SO(4)\otimes\SO(4)\\
&&\SO(6)\otimes\SO(6)\otimes\SO(4)\\
&&\SO(6)\otimes\SO(4)
\end{eqnarray*}
respectively.

In the same way, from a generic superpotential in the $16$-plets,
$W(\psi,\bar\psi)$, one obtains for them a vacuum solution which preserves
the relative~SU(5)'s, with expectation values, for the different
$\md{\psi_i}=\md{\bar{\psi}_i}$, which either vanish or coincide
(for a superpotential including up to quartic terms).
As in standard~SO(10), with the appropriate embedding of the~SU(5)'s
in the corresponding~SO(10)'s, the intersection of the different
subgroups leads to the residual $\SU(3)\otimes\SU(2)\otimes{\rm U}(1)$.
Of course, to avoid uneaten Goldstone-bosons, the superpotential must also
contain some interaction terms between the $Z$ and the $\psi,\bar{\psi}$
multiplets. The terms of lowest dimensionality are
\begin{eqnarray*}
W(Z,\psi,\bar{\psi})  &=&  \frac{1}{M^2}\sum_{i,k}Z_{a_ia_k}\big\{ \lambda
(\psi\gamma_{a_i}\psi)(\psi\gamma_{a_k}\psi)+\\
&&+\sigma
(\bar{\psi}\gamma_{a_i}\bar{\psi})(\psi\gamma_{a_k}\psi)+\tau
(\bar{\psi}\gamma_{a_i}\bar{\psi})(\bar{\psi}\gamma_{a_k}\bar{\psi})
\big\}\end{eqnarray*}
Such terms act as connectors between the $Z$ and the $\psi,\bar{\psi}$ fields
in the sense that they vanish on the vacuum configurations described
above (the~10 of SO(10) has no singlet under SU(5)) and they give masses
to all unwanted Goldstone bosons.

In analogy with the~SU(5) case, to solve the proton decay problem,
the triplets of Higgs fields entering into the Yukawa superpotential
must occur in two replicas of 10-plets, $H_i,H'_i$ and be coupled
to a pair of $Z$-type triplets
\begin{equation}\label{eq:WHZ10}
W(H,Z)=HZH+H'Z'H'.
\end{equation}
With an appropriate choice of the $Z,Z'$ vacuum expectation values, one gives
masses to all coloured triplets in $H,H'$ and remains with a pair of
SU(2)-doublets in $H_3$ ($h_3$ and $\bar{h}_3$) and a pair of
doublets in $H'_2$ ($h'_2$ and $\bar{h}'_2$).
Since $h_3$ is coupled to the up quarks (and the neutrinos), whereas
$\bar{h}'_2$ gives mass to the down quarks and the charged leptons,
in order not to undo the successful unification of couplings,
the two remaining doublets have to receive a heavy mass.
This is achieved by the couplings, asymmetric in $\psi,\bar{\psi}$,
\begin{equation}\label{eq:WHy}
W(\psi,\bar{\psi},H)=H\bar{\psi}\bar{\psi}+H'\psi\psi.
\end{equation}

The Yukawa superpotential with the $16$-plets of matter families,
always omitting dimensionless couplings and contractions over SO(10) and
permutation indices, is
\begin{eqnarray}\label{eq:WY10}
W_Y &=& \sum_n\frac{1}{M_u^n} H Y^n (16\,16)+\sum_{n'}
\frac{1}{M_d^{n'+1}}  H'Y'Y^{n'}(16\,16)+\\
&&+\sum_{n,m}\frac{1}{M_u^{n+m+2}}  HY^n(16\,\psi) Y^m(\psi\,16)+
\sum_{n',m'}\frac{1}{M_d^{n'+m'+3}}H'Y'Y^{n'}(16\,\psi)Y^{m'}(\psi\,16)
\nonumber\end{eqnarray}
The first two terms, in conjunction with~(\ref{eq:WHZ10}) and~(\ref{eq:WHy}),
give rise to the diagonal terms in $M^{\rm u}$ and $M^{\rm d},M^{\rm e}$
respectively, whereas the last two are the source of the off diagonal
matrix elements.
Notice in particular in these last terms the necessity to introduce
the $\psi$-fields.

After insertion of the various vacuum expectation values, this
superpotential gives rise to the same mass matrices as
in~(\ref{sys:ude}a,b,c), where now
\begin{equation}
t_d = \frac{\md{\psi}^2\md{T_Y}}{M_d^3},\qquad
t_u = \frac{\md{\psi}^2\md{T_Y}^2}{M_u^4},
\end{equation}
whereas the $\varepsilon$ parameters keep the same meaning as before.
As a consequence, the description of all charged fermion masses and
mixings is unchanged.

Needless to say, a special phenomenological interest of~SO(10),
or~$\SO(10)^3$, is that it allows a discussion of neutrino masses.
{}From the superpotential~(\ref{eq:WY10}), the only relevant neutrino mass
terms are the diagonal entries in the Dirac $\nu\nu^c$ mass matrices,
which are effectively
identical to the $Q=2/3$ quark masses, since the off-diagonal
terms are too small in both cases.
Such diagonal entries arise from the first term in~(\ref{eq:WY10}),
which in any case respects the Pati-Salam~SU(4).

An independent source for Dirac masses, $\nu\nu^c$, and only for them,
is the operator
$$\frac{1}{M_\nu^{n+2}}\sum_k(16_k\bar{\psi}_k)\cdot
 \sum_i (H Y^n)_{a_i}(16_i\gamma_{a_i}\psi_i)$$
which, depending on $M_\nu$, could give significant contributions to diagonal
and off diagonal entries.
As mentioned, we assume that such factorized operators are suppressed.

Always disregarding factorized operators, the dominant contribution to the
Majorana mass term $\nu^c\nu^c$ for the right-handed neutrinos comes from the
dimension-4 superpotential term
$$\frac{1}{M_\nu}\sum_i
(\bar{\psi}_i\Gamma_{a_i}^{(\overline{126})}\bar{\psi}_i)
(\psi_i\Gamma_{a_i}^{(126)}\psi_i)$$
(or equivalent, by SO(10) Fiertz rearrangement),
where, inside each parenthesis, the spinorial indices are contracted
to make a~126 ($\overline{126}$) representation of~SO(10).
Such term, taking $\md{\psi_i}$ $i$-independent as explained above,
gives a right handed neutrino mass matrix proportional to the identity
by a factor $\md{\psi}^2/M_\nu$.
On the other hand, the dominant non factorizable contribution to a direct
left-handed neutrino Majorana mass, $\nu\nu$, comes from an operator of
very high dimension and can therefore be neglected.
As a overall consequence, the effective $\nu\nu$ mass matrix,
in the original basis, is, to a good approximation, diagonal, with eigenvalues
\begin{equation}
m_{\nu_\tau}=\frac{m_t^2}{\md{\psi}^2/M_\nu},\qquad
m_{\nu_\mu}=m_{\nu_\tau}\frac{m_c^2}{m_t^2},\qquad
m_{\nu_e}=m_{\nu_\mu}\frac{m_u^2}{m_c^2},
\end{equation}
where the $Q=2/3$ quark masses refer to their high scale values.
Taking $M_\nu \le 10^{19}\,{\rm GeV}$ and $\md{\psi}\approx 10^{16}\,{\rm
GeV}$,
we have, for $\lambda_t^{\rm GUT}=0.3\div4$
\begin{equation}\label{eq:mnu}
m_{\nu_\tau}\le 100\,{\rm eV},\qquad
m_{\nu_\mu}=(0.05\div3)10^{-5}\, m_{\nu_\tau},\qquad
m_{\nu_e}  =(0.6 \div4)10^{-5}\, m_{\nu_\mu}.
\end{equation}
As seen in section~5, in the original basis,
the charged lepton mass matrix~(\ref{sys:ude}c) is not flavour diagonal.
As a consequence, the mixing angles in the leptonic charged current weak
interactions arise from the need to diagonalize $M^{\rm e}$, so that
\begin{equation}\label{eq:Vnu}
\begin{array}{l}
\sin2\theta_{e\mu}\approx\sin2\theta_{\mu\tau}\approx 2\ed\approx0.1,\cr
\sin2\theta_{e\tau}\approx2\ed^2\approx0.5\cdot10^{-2}
\end{array}
\end{equation}

It is clear from eq.s~(\ref{eq:mnu},\ref{eq:Vnu}) that a
$\tau$-neutrino saturating the bound~(\ref{eq:mnu}) could give, depending
on the value of $\lambda_t^{\rm GUT}$, a
consistent solution of the solar neutrino problem as a manifestation
of a $\nu_e$-$\nu_\mu$ resonant~\cite{nuOsc} oscillation.
In such a case the $\nu_\mu$-$\nu_\tau$ oscillation would
most likely give an observable signal
in the forthcoming experiments at {\sc CERN}~\cite{CHORUS}.

\section{Conclusions}
In conclusion, we have shown how a realistic pattern of fermion masses
and mixings can be obtained in a unified supersymmetric theory
where flavour is treated in a completely symmetric way.
Although we do not obtain any particular relation between different
masses and/or mixings angles of the charged fermions,
due to the presence of unknown coefficients
of order unity, we do find a simple rationale for these masses and mixings
in terms of~3 large (grand unified) mass scales.
Some particular relations might occur as a consequence of the choice
of a more specific vacuum structure than the one that we have considered.

By going to~SO(10), we predict the three neutrino masses in terms of an
overall scale and, within factors of order unity, their mixing angles.
A $\nu_e$-$\nu_\mu$ oscillation is suggested as a solution of the solar
neutrino problem, implying a visible $\nu_\mu$-$\nu_\tau$ oscillation
in the forthcoming experiments.

We have been independently laid to consider unified models of the type
discussed on
this paper by requiring that the unified gauge group be broken by the
vacuum expectation values of Higgs fields in the fundamental representation.
This criterium is suggested by string theory considerations.
Models with the gauge structure and with the Higgs content suggested here
seem indeed to be obtainable in a string theory context~\cite{StringAndGUT}.
Also based on these considerations, we suggest that~$\SO(10)^3$ is the
largest possible `sensible' subgroup of~SU(45) (or rather~SU(48))
that can be gauged
and can be broken to the standard $\SU(3)\otimes\SU(2)\otimes{\rm U}(1)$ group
by Higgs fields in simple enough representations.
At the same time, we find it particularly interesting that the
residual global symmetry, in absence of Yukawa couplings,
does not contain any non-abelian flavour symmetry.


\frenchspacing
\begin{thebibliography}{22}
\bibitem{GUTAndString} R. Barbieri, G.Dvali and A. Strumia,
Pisa preprint IFUP -- TH 20/94, to appear in Phys. Lett. {\bf B}.
\bibitem{StringAndGUT} L. Ib\`a\~nez,
talk given at Susy 94, Ann Arbor, Michigan, May 1994.
\bibitem{Strings} For a review and references, see
M. B. Green, J. H. Schwarz and E. Witten, ``{\em Superstring theory\/}'',
Cambridge monographs on mathematical physics, Cambridge, 1986.
\bibitem{Froggatt} C. Froggatt, G. Lowe and H. Nielsen,
Phys. Lett. {\bf B311} (1993) 163 and references therein.
\bibitem{Dvali} G. Dvali, Phys. Lett. {\bf B287} (1992) 101.
\bibitem{RGE} See, e.g., V. Barger, M. Berger and P. Ohmann,
Phys. Rev. {\bf D47} (1993) 1093 and references therein.
\bibitem{nuOsc} L. Wolfenstein,
Phys. Rev. {\bf D17} (1978) 2368; {\bf D20} (1979) 2634;\\
S. Mikheyev and A. Smirnov, Yad. Fiz. {\bf 42} (1985) 1441;
Nuovo Cim. {\bf 9C} (1986) 17.
\bibitem{CHORUS}
{\sc CHORUS} Coll., N. Armenise et al., {\sc CERN--SPSC}/90-42 (1990);\\
{\sc NOMAD} Coll., P. Astier et al., {\sc CERN--SPSC}/91-21 (1991).
\end{thebibliography}
\end{document}